\documentstyle[12pt, A4,epsfig]{article}
\thicklines                     
\begin{document}
\begin{titlepage}
\begin{centering}
\vspace{2cm}
{\Large\bf Domain walls, stabilities, and the mass hierarchy}\\
\vspace{0.5cm}
{\Large\bf of the Randall-Sundrum Model}\\
\vspace{2cm}
Haewon~Lee\footnote{E-mail address: {\tt hwlee@cbucc.chungbuk.ac.kr}} and
W.~S.~l'Yi\footnote{E-mail address: {\tt wslyi@cbucc.chungbuk.ac.kr}}
\vspace{0.5cm}

{\em Department of Physics}\\
{\em Chungbuk National University}\\
{\em Cheongju, Chungbuk 361-763, Korea}\\
\vspace{1cm}
\begin{abstract}
\noindent
Randall-Sundrum model, which has a scalar field, is used to investigate
the domain structure of the extra dimension and to obtain a possible solution of
the mass hierarchy problem.
It is found that when the domain wall size is comparable to that of domains,
domains become unstable.
To construct a reliable theory, a region of physical parameter space, where domains are stable,
is identified.
Analytic forms of field configurations are obtained by perturbative expansions
in term of a small parameter that is approximately equal to the relative size of domain wall
with respect to domains.  By placing a single 3-brane in one of the domain,
one can solve the mass hierarchy problem.
\end{abstract}
\end{centering}
\vspace{.5cm}
\vfill

Nov.~15, 2000\\
\end{titlepage}

\setcounter{footnote}{0}
Randall-Sundrum model\cite{RS1} inspired a novel way to the solution of the
old mass hierarchy problem, that is, why are the masses of ordinary particles
almost negligible compared to the Planck mass?  The main idea is that by
introducing two worlds\cite{HW}, one hidden and another visible, separated by a small distance
along an extra dimension, may produce exponentially large Planck mass when branes of two worlds
are allowed to interact through the Einstein gravity.
In their sequel paper\cite{RS2} they have
shown that the invisible hidden world may be infinitely separated from the visible world,
and then the five dimensional gravity interaction produces a single normalizable bound
state mode that propagates along the brane where the visible world is\cite{RS2,local,Chaichian}.
The continuum modes of massive Kaluza-Klein states contribute a testable correction to the
Newtonian limit\cite{GF,Newton,Duff,Gregory}.

After their pioneering works, investigations along various directions
were done during the last few years.  The detailed analysis of the gravitational fluctuations
of the Randall-Sundrum vacuum configuration is one of the next imperative steps\cite{GF,GF2,Gar}.
Another is the inclusion of more than two branes, either thin or thick ones\cite{MB,MB2}.
Still another possibility is the introduction of a scalar field which acts as an
initiating field of the domain structure\cite{scalar,SF,SF2}.  Ichinose\cite{ichinose} found
interesting results that there exists an exact solution of the field
configuration when one assumes a series expansion in terms of hyperbolic tangent functions,
and that the spacetime has the Randall-Sundrum type vacuum.

In this paper we ask questions such as, ``Is it possible to describe
the Randall-Sundrum scenario with a single brane, and if it is the case what is the
physical meaning of the location of the brane in the extra dimension?
Are these field configurations stable?  If it is the case, is there any restriction on the
values of physical parameters?''
To answer this question we make use of a five dimensional Randall-Sundrum type model
with a scalar field which Ichinose used.
The scalar field, which is different from the Higgs field of the brane world,
is introduced to determine the domain structure along the extra dimension.
It can be shown that when the flow lines corresponding to the evolutions of phase points
dictated by the equations of motion are drawn in a certain phase space,
there is a clear-cut region of a physical parameter space where the domain structure
is unstable.  One of the two controlling parameters which is critical to the stability is
$\epsilon$ which is a measure of the relative size of the domain wall with respect
to domains.  If the size of domain wall is comparable to that of domains,
domains become unstable and tend to squeeze down to the domain wall.
But for a relatively narrow domain wall, domains remain stable.
In this case it is possible to solve the equations of motion
analytically when the usual perturbative expansion in terms of $\epsilon$ is performed.
The analytic solutions clearly show the domain structure.
When the origin of the extra dimension is chosen to be at the center of the domain wall,
the spacetime metric has the correct $AdS_5$ geometry in the domain,
and is almost flat at the domain wall.  If the relative size
of domain wall with respect to domains goes to zero, it reproduces the exact Randall-Sundrum
vacuum geometry.  By placing a single 3-brane at $y_b,$ a location in
one of the domains, the Randall-Sundrum type solution of the mass hierarchy problem may be
reproduced.  The extra coordinate $y_b$ effectively acts as the ``compactification
radius.''   Furthermore there is no need of fine-tuning of the cosmological
constants of the visible and hidden worlds that is needed in the original
Randall-Sundrum model.

In this paper, the equations of the motion of the model are presented in section 1.
By introducing a phase space, stabilities and a region of the parameter
space where domains are stable are analyzed in section 2.
The perturbative analytical solutions of fields expanded in terms of $\epsilon$ are
given in section 3. Domain structures and the mass hierarchy problem are consulted
in the last section.
\section{Action and the equations of motion}
We consider a $D$-dimensional Randall-Sundrum model which
has the spacetime coordinate patch given by $x^A = (x^\mu,y),$ $\mu=0,...,D-2,$
where $y=x^{D-1}$ is the coordinate of the extra
dimension which may take any value in $-\infty < y < \infty.$
The action we are considering is given by
\begin{equation}\label{S}
S =\int d^{D}x \sqrt{-g} \left[ \delta(y-y_b)L_{matter} +
  \Lambda -{1\over 2}M^{D-2} R_{D}
  -{1\over 2}g^{AB}\partial_A\Phi\partial_B\Phi - V(\Phi)\right]
\end{equation}
where $L_{matter}$ is a Lagrangian of matter fields which reside in a
$(D-2)$-brane located at $y=y_b,$ and
$M$ is the $D$-dimensional Planck mass which one should introduce to leave
the action dimensionless.  The detailed form of matter Lagrangian is irrelevant
for determining the vacuum configuration of graviton,
and we assume $L_{matter}=0.$
$R_{D}$ is the $D$-dimensional curvature scalar, and
the cosmological constant, for the convenience of later calculations, is taken to be
$-\Lambda.$    For the scalar potential $V(\Phi)$
and metric tensor $g_{AB},$ we assume the usual forms
\begin{eqnarray}
V(\Phi)&=&\lambda (\Phi^2-\upsilon^2)^2, \\
ds^2_{D}&=&{\rm e}^{-2\sigma(y)}ds^2_{D-1} +dy^2.\label{ansatz}
\end{eqnarray}
Here $ds^2_{D-1}=\eta_{\mu\nu}(x)dx^\mu dx^\nu$ is a
$(D-1)$-dimensional metric tensor such that
$ds^2_{D-1} > 0$ for spacelike intervals.
If the ansatz (\ref{ansatz}) is truly correct, it can be used in the calculation
of the action.  In fact, the usual Einstein's field equations,
derived from (\ref{S}), can be compared to the equations obtained from the Lagrangian
using the ansatz first. But the results are the same.  For the convenience of calculations
we use the later algorithm to obtain the vacuum geometry.

The $D$-dimensional curvature scalar written in terms of
the $(D-1)$-dimensional one is given by
\begin{equation}
R_{D}={\textstyle e}^{2\sigma(y)}R_{D-1}
-2(D-1)\left({\partial^2 \sigma \over \partial y^2}
- {D\over 2}\left({\partial \sigma\over \partial y}\right)^{2}\right).
\end{equation}
The Lagrangian corresponding to (\ref{S}) is given by
\begin{eqnarray}\label{lagrangian}
{\cal L}&=& -{1\over 2}e^{-(D-3)\sigma}
   \left[ M^{D-2}R_{D-1} +\eta^{\mu\nu}\partial_\mu\Phi\partial_\nu\Phi\right]
  \\ \nonumber
   &&-e^{-(D-1)\sigma}\left[
    -\Lambda +  {1\over 2}\left({\partial \Phi \over \partial y}\right)^2
      + \lambda(\Phi^2-\upsilon^2)^2
      -{1\over 2}(D-1)(D-2)M^{D-2}
      \left({\partial \sigma\over \partial y}\right)^{2}\right].
\end{eqnarray}
Since we are interested in the domain structure, we assume that $\Phi$
depends only on $y$\cite{H}.
The equations of motion corresponding to the
Lagrangian (\ref{lagrangian}) are
\begin{eqnarray}
{1\over 2}M^{D-2}e^{2\sigma}R_{D-1}
 &=&  {1\over 2}\left({d\Phi \over dy}\right)^2
 - \lambda(\Phi^2-\upsilon^2)^2 + \Lambda
 - {(D-1)(D-2) \over 2}M^{D-2}\left( {d\sigma \over dy}\right)^2, \\
\left({d\Phi \over dy}\right)^2
 &=& (D-2)M^{D-2} {d^2\sigma \over dy^2}
+ {1 \over D-1} M^{D-2} e^{2\sigma}R_{D-1},\\
R_{D-1}^{\mu\nu} &=& {1\over D-1}\eta^{\mu\nu}R_{D-1},
\end{eqnarray}
where $R_{D-1}$ is constant.

In this paper, we restrict our consideration only to $D=5$ case which has the
flat four dimensional metric given by $\eta_{\mu\nu}={\rm diag}(-1,1,1,1).$
Then the equations of motion reduce to
\begin{eqnarray}
6M^3\left({d \sigma \over d y}\right)^2
&=& {1\over 2}\left({d \Phi \over d y}\right)^2
- \lambda(\Phi^2-\upsilon^2)^2 +\Lambda,\\
3M^3 {d^2 \sigma \over d y^2} &=& \left({d \Phi \over d y}\right)^2.
\end{eqnarray}
To pay our attention to the essence of the equations, we introduce dimensionless
variables such as
\begin{eqnarray}
\varphi &=& \upsilon^{-1}\Phi, \\
z &=& \sqrt{ 3\Lambda M^3 \over 2 \upsilon^4} y.
\end{eqnarray}
Even though $\sigma$ is dimensionless, it is useful to define new dimensionless
variables $s$ and $\zeta$ given by
\begin{eqnarray}
s &=& {3M^3 \over \upsilon^2}\sigma,\\
\zeta &=& {ds \over dz}.
\end{eqnarray}
The equations of motion, when written in terms of these, are
\begin{eqnarray}
\zeta' &=& (\varphi')^2, \label{em}\\
\zeta^2 &=& 1 + {3M^3 \over 4\upsilon^2} \varphi'^2
 - {\lambda \upsilon^4 \over\Lambda}(\varphi^2-1)^2,\label{em2}
\end{eqnarray}
where the symbol $'$ denotes the differentiation with respect to $z.$
These are the basic equations that we begin with to discuss stabilities
in the next section.
\section{Stability analysis of the Randall-Sundrum model}
The basic equations (\ref{em}) and (\ref{em2}) we derived in the last section are
nonlinear coupled differential equations that do not allow us to obtain
analytic solutions in closed forms.
But fortunately these are first order differential equations whose general behavior
can be visualized in a phase space constructed by $(\varphi,\zeta).$
Flows of phase points are determined by
\begin{eqnarray}\label{dzeta/dphi}
{d \zeta \over d \varphi} &=& {\zeta' \over \varphi'} = \varphi' \nonumber \\
&=& \pm A \sqrt{(1-\varphi^2)^2 - \epsilon(1-\zeta^2)}, \label{flow_eq}
\end{eqnarray}
where the dimensionless parameters $A$ and $\epsilon$ are given by
\begin{eqnarray}
A &=& \sqrt{ 4\lambda \upsilon^6 \over 3\Lambda M^3}, \\
\epsilon &=& {\Lambda \over \lambda\upsilon^4}.
\end{eqnarray}
It is sufficient to consider only the branch corresponding to the positive
sign in (\ref{dzeta/dphi}).
Since ${d\zeta / d\varphi}$ is a real function, it is clear that the allowed
region of the phase space is restricted by the condition
$C_\epsilon(\varphi,\zeta)\geq 0,$ where
\begin{equation}
C_\epsilon(\varphi,\zeta)=(1-\varphi^2)^2 - \epsilon(1-\zeta^2).
\end{equation}
The shape of the boundary of the forbidden region, which we name it
the island, depends on $\epsilon.$
The islands for various values of $\epsilon$ are drawn in Fig.1.
\begin{figure}
\begin{center}
\psfig{figure=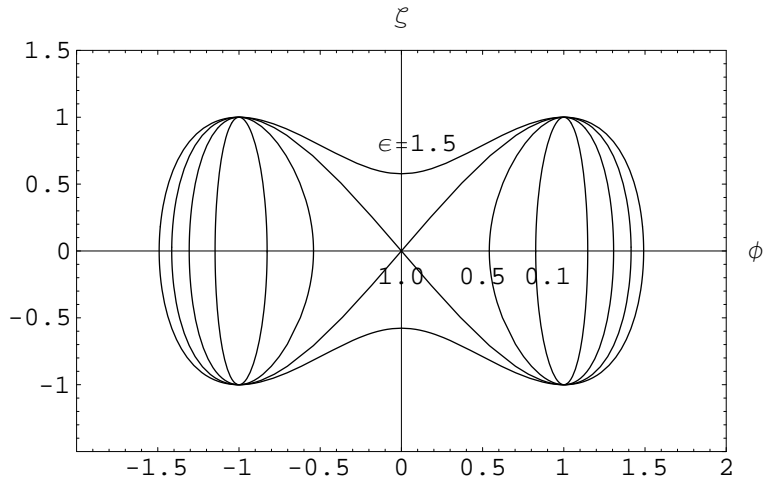}
\end{center}
\caption{The shapes of boundaries of the forbidden regions depend only on
$\epsilon.$  For $\epsilon<1,$ it consists of two disjoint islands. When $\epsilon=1,$
it touches the origin of the phase space, but the origin is still allowed.
If $\epsilon> 1,$ the two forbidden islands merge to one big island.
In this case, the origin is not allowed}
\end{figure}
They are symmetric under the separate reflections of $\varphi$ and $\zeta.$
When we assume that the domain wall is located at  $z=0,$ it becomes clear that
the origin of the phase space, $\varphi=\zeta=0,$ should be
in the allowed region of the phase space.
It is possible only when $0 < \epsilon\leq 1.$

The coefficient $A$ given in (\ref{dzeta/dphi}) determines the initial flow
direction at the origin $\varphi=\zeta=0.$
The same equation shows that for a given shape of island, which
is determined by $\epsilon,$ flows in the phase space either
terminate at the island or diverge indefinitely depending on $A.$
There is unique $A(\epsilon)$ by which the flow line starting at the origin
terminates at $\varphi=\zeta= 1.$
When $A$ is less than this critical value $A(\epsilon),$
the flow line reaches the island, and stays there forever. But if it is slightly larger than
$A(\epsilon),$ the flow bypasses the island, and runs indefinitely.
In this case it becomes unstable.
The numerical result of computation of this behaviour is plotted in Fig.2.
\begin{figure}
\begin{center}
\psfig{figure=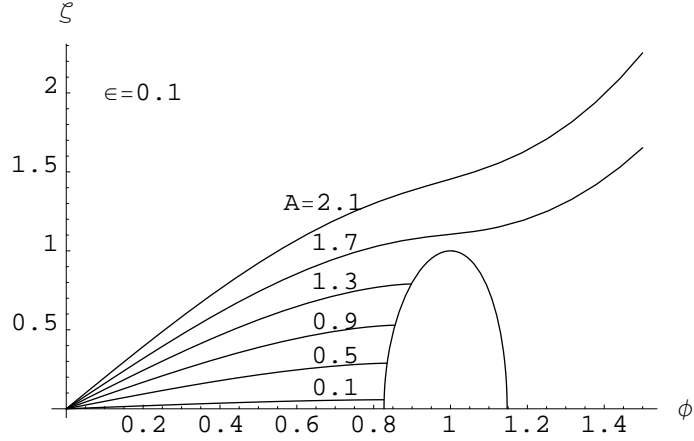}
\end{center}
\caption{Flows of phase points terminate on the boundary of the forbidden
region, that is the island, when $A\leq A(\epsilon).$
If $A=A(\epsilon),$ it terminates at the critical point $\varphi=\zeta=1.$
But if $A>A(\epsilon),$ $\zeta(\varphi)$ grows indefinitely.}
\end{figure}
To understand stabilities one should
find $A(\epsilon)$ corresponding to the critical flow.
This is investigated in the next section where analytical solutions of field configurations
are also found.
%
\section{Perturbative analytical solutions of fields}
Even though the flow equation is highly nonlinear, one may use the usual
perturbation technique to solve it.
We assume that $\varphi$ and $\zeta$ both reach the critical values,
$\varphi\to 1$ and $\zeta\to 1,$ as $z \rightarrow \infty.$
We have seen, for the stable solution, that $0< \epsilon\leq 1.$
It allows us to expand the flow equation (\ref{dzeta/dphi})
in terms of $\epsilon,$
\begin{equation}\label{pert_flow_eq}
{d \zeta \over d \varphi} = A [
1-\varphi^2 -{\epsilon \over 2}{1-\zeta^2 \over 1 - \varphi^2}
+ {\epsilon^2 \over 8}{(1-\zeta^2)^2 \over (1 - \varphi^2)^3}
+\cdots].
\end{equation}
One can solve this equation perturbatively under the conditions
\begin{equation}
\zeta(\varphi=0) = 0, \;\;\; \zeta(\varphi=1) = 1.
\end{equation}

Firstly, we find $A(\epsilon)$ corresponding this critical flow,
and then solve (\ref{pert_flow_eq}).
The curve $A(\epsilon)$ in the parameter space $(\epsilon, A)$ divides it up
into the stable and unstable regions.
The general formula for $A(\epsilon)$ is hidden in
\begin{equation}\label{genA(e)}
\int^{1}_0 {d\zeta \over d\varphi}d\varphi = 1.
\end{equation}
Using the $\epsilon$ independent part of (\ref{pert_flow_eq}), we have
\begin{equation}
1 = A(\epsilon) \int^1_0 (1-\varphi^2)d\varphi \\ \nonumber
   = {2\over 3}A(\epsilon).
\end{equation}
That is, to the order of ${\cal O}(\epsilon^0),$ $A(\epsilon)={3\over 2}.$
Then, by (\ref{pert_flow_eq}), one has
\begin{equation}\label{zeta_0}
\zeta={3\over 2}(\varphi -{1\over 3}\varphi^3).
\end{equation}
Substituting this in (\ref{pert_flow_eq}) again, one gets the following equation,
\begin{equation}\label{O(1)}
{d \zeta \over d \varphi} = A(\epsilon)
[  1-\varphi^2 -{\epsilon \over 2}{1-{9\over 4}
(\varphi-{\varphi^3\over 3})^2
 \over 1 - \varphi^2} ].
\end{equation}
>From this we find that $A(\epsilon)$ and $\zeta(\varphi),$
to the order ${\cal O}(\epsilon),$ are
\begin{eqnarray}
A(\epsilon) &=& {{3\over 2}\over 1- {19\over 40}\epsilon},\\
\zeta(\varphi) &=& \zeta_0(\varphi) + \epsilon\zeta_1(\varphi).\label{zeta(phi)}
\end{eqnarray}
Here $\zeta_0(\varphi)$ is the same as (\ref{zeta_0}), and
\begin{equation}
\zeta_1(\varphi)=-{3\over 80}(\varphi - 2\varphi^3 +\varphi^5).
\end{equation}

To solve $\varphi$ as a function of $z,$  we combine
(\ref{dzeta/dphi}) and (\ref{pert_flow_eq}),
\begin{equation}
{d\varphi \over dz}= {d\zeta\over d\varphi}
= A(\epsilon)
[
 1-\varphi^2 -{\epsilon \over 2}{1-{9\over 4}
(\varphi-{\varphi^3\over 3})^2
 \over 1 - \varphi^2}
].
\end{equation}
This can be integrated to give the following
\begin{equation}\label{z(phi)}
z={\epsilon\over 12}\varphi + {10-\epsilon\over 30}\log{1+\varphi\over 1-\varphi},
\end{equation}
It is valid up to the order ${\cal O}(\epsilon).$
Solving $\varphi$ in terms of $z,$ one gets
\begin{equation}\label{v_z}
\varphi(z) =\tanh {3z\over 2} +{3\over 2}\epsilon
\left( {z\over 10} - {1\over 12}\tanh{3z\over 2}\right)
(1-\tanh^2 {3z\over 2}) + {\cal O}(\epsilon^2).
\end{equation}
Note that there is a $z$-linear term which is not easy to
guess\cite{ichinose}.
\begin{figure}
\begin{center}
\psfig{figure=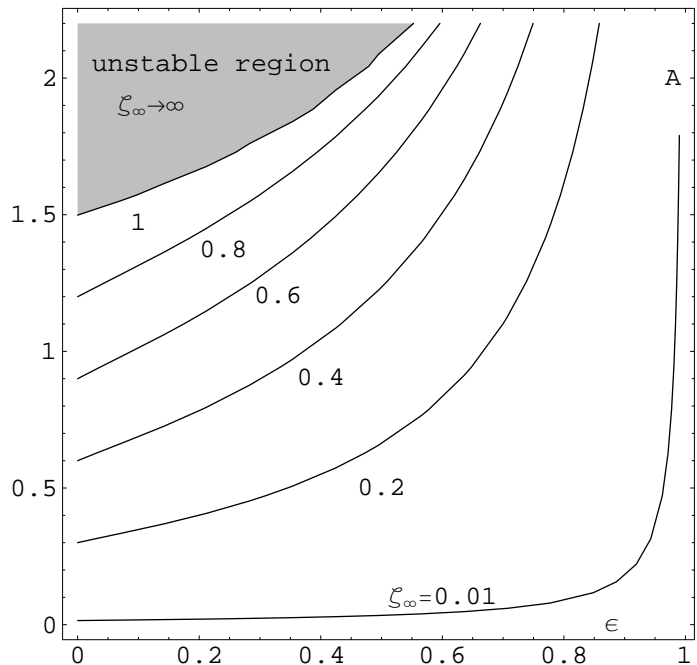}
\end{center}
\caption{The fate of field configuration depends on $\epsilon$ and
$A.$ At the dark region of the phase space the field configuration
is unstable, and $\zeta$ runs indefinitely as $z$ becomes larger.
The boundary of this unstable region is described by the curve
$A=A(\epsilon).$
 If $A\leq A(\epsilon),$ $\zeta$ has the limiting value
$\zeta_\infty$ as $z\to\infty.$ For a given $\epsilon,$ there is
unique $A = A(\epsilon,\zeta_\infty)$ by which flows of phase
points determined by (\ref{dzeta/dphi}) terminate at
$\zeta_\infty.$ Various curves correspond to various
$\zeta_\infty.$ Note that $A(\epsilon) = A(\epsilon,1).$}
\end{figure}
When (\ref{v_z}) is used in (\ref{zeta(phi)}), one may have $\zeta=\zeta(z).$
The corresponding graph is plotted in Fig.4.
\begin{figure}
\begin{center}
\psfig{figure=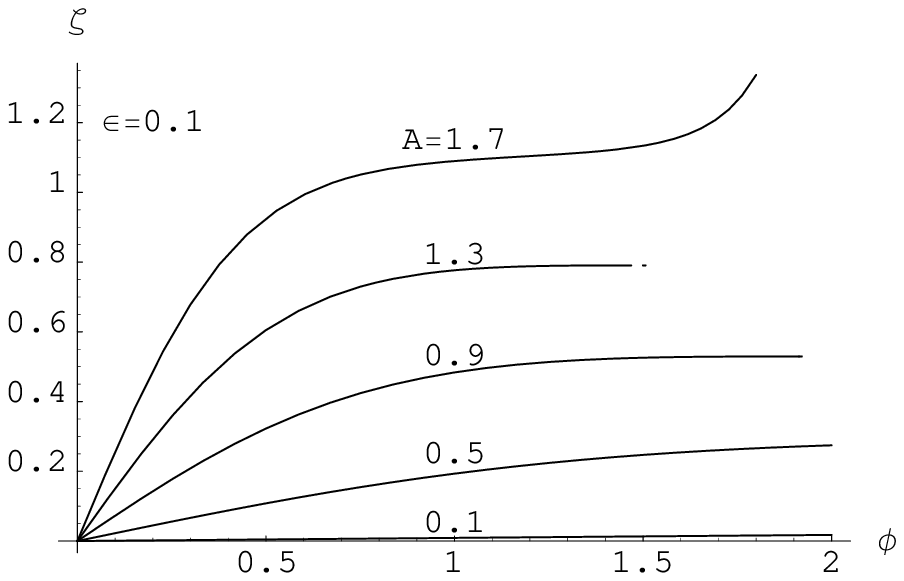}
\end{center}
\caption{For a given $\epsilon,$ $\zeta(z)$ saturates if $A\leq A(\epsilon).$
When $A= A(\epsilon),$ $\zeta(z)$ saturates exactly to 1.}
\end{figure}
It is clear that both $\varphi$ and $\zeta$ approach the domain values quickly.
Comparing the result with the original Randall-Sundrum model\cite{RS1},
it shows that the domain structure of $\varphi$ smoothes out the singular behavior
of $\zeta$ at the origin.
\section{Domain structures and the mass hierarchy problem}
The original Randall-Sundrum model is a theory where the size of domain wall is zero.
In this paper the domains and domain wall have finite sizes whose
analytical structures are determined by the
relation (\ref{v_z}).  Using this equation we determine the domain wall size
$\mu$ and the domain size $\xi.$
According to (\ref{lagrangian}), an observer who lives in the 3-brane located
at $y=y_b$ effectively measures the extra dimension $y$ with an effective scale factor
$e^{-(D-3)\sigma(y)},$ where $D=5.$
This means that
\begin{equation}
\int^\infty_{-\infty} e^{-2\sigma(y)}dy = \mu + 2\xi.
\end{equation}

To compute the domain wall size, we use ({\ref{em}) and (\ref{em2}). Fig.4 shows that
$\zeta(z)$ is nearly constant, $\zeta_\infty= \zeta(z\to\infty)$ or
$-\zeta_\infty,$ in the domains, and almost linear at the transitory domain wall.
Here, $\zeta_\infty <1$ for a stable configuration.
Both $\varphi$ and $\zeta$ vanish at the center of the domain wall. That is,
the slop of $\zeta(z)$-curve at the origin is given by
\begin{equation}
\zeta'_0 = \varphi'^{2}_0 = {4\upsilon^2\over 3M^3}\left({1\over \epsilon }-1\right).
\end{equation}
The $z$-coordinate $\delta$ of the boundary of the domain wall is given by
\begin{equation}
\delta={\zeta_\infty\over \zeta'_0}
= {3M^3\over 4\upsilon^2}{\epsilon\zeta_\infty\over 1-\epsilon}.
\end{equation}
For $\epsilon\approx 0,$ it can be simplified to
\begin{equation}
\delta\approx{3M^3\over 4\upsilon^2}\epsilon\zeta_\infty.
\end{equation}
Note that $z$ is the physical coordinate
$y$ measured in the unit of $\ell,$ where
\begin{equation}
\ell = \sqrt{2\upsilon^4 \over 3\Lambda M^3}.
\end{equation}
We approximate $\zeta(z)={ds/dz}$ in the following way,
\begin{equation}
{ds\over dz} =
\left\{
  \begin{array}{ll}
    \zeta_\infty & \mbox{if $z > \delta,$} \\
   \zeta_\infty{1\over\delta}z & \mbox{if $-\delta < z < \delta,$} \\
   -\zeta_\infty & \mbox{if $ z < -\delta.$} \\
  \end{array}
\right.
\end{equation}
This can be solved to give
\begin{equation}
s(z) =
\left\{
  \begin{array}{ll}
   \left({1\over2\delta}z^2 +{1\over2}\delta\right)\zeta_\infty
                   & \mbox{if $-\delta < z < \delta,$} \\
   \zeta_\infty|z| & \mbox{otherwise,}
  \end{array}
\right.
\end{equation}
where ${1\over2}\delta$ is the integration constant which is needed to leave $s(z)$
both continuous and smooth at the boundary of the domain wall, $z=\pm\delta.$

The domain wall size is given by
\begin{equation}
\mu = \ell
    \int^{\delta}_{-\delta}
      e^{ - 2( { z^2\over 2\delta} + {\delta \over 2})\zeta_\infty{\upsilon^2 \over 3M^3}}dz
            \approx {2\delta\ell},
\end{equation}
where the last approximation is valid when $\epsilon\approx 0.$
On the other hand, the domain size $\xi$ is give by
\begin{equation}
\xi = \ell
\int^\infty_\delta e^{-{2\upsilon^2\over3M^3}\zeta_\infty z}dz
= {2\delta\ell\over \zeta^2_\infty} {1-\epsilon\over \epsilon}
 e^{-{\epsilon\zeta^2_\infty\over 2(1-\epsilon)}},
\end{equation}
which, for $\epsilon\approx 0,$ can be reduced to the following simple form
\begin{equation}
\xi\approx{1\over \epsilon\zeta^2_\infty}\mu.
\end{equation}

When $\epsilon\to 1,$ the above equation states that the sizes of domain and
domain wall are comparable implying that our approximation is not valid.
For unstable field configurations such as $A > A(\epsilon),$ both $\sigma(y)$ and $\Phi(y)$ grow
enormously.  These mean that the potential energy density is very
large in the domain. But the effective scale factor ${\textstyle e}^{-4\sigma(y)}$
vanishes rapidly, thus pushing domains to the domain wall.
This sustains the total potential energy still to a finite value.

Now consider a possible solution of the mass hierarchy problem.  The Planck mass $M_{Pl},$
inferred from the Lagrangian (\ref{lagrangian}), is given by
\begin{equation}
M^2_{Pl} = M^3\int^\infty_{-\infty} e^{-2\sigma(y)}dy,
\end{equation}
which, for a relatively narrow domain wall, is equal to
$M^2_{Pl} \approx {1\over \epsilon\zeta^2_\infty}\mu M^3.$
To compare physical mass with the Planck mass, we place a 3-brane at the coordinate
$y_b = \ell z_b,$  and then turn on the interactions among particles which live in the brane.
The usual form of the action of a Higgs field is the following,
\begin{equation}
S_H = -\int d^4 x \sqrt{-\bar{g}}\left[\bar{g}^{\mu\nu}D_\mu H^\dagger D_\nu H
-\lambda_H(|H|^2 - \upsilon_H^2)^2\right],
\end{equation}
where ${\bar g}_{\mu\nu}(x) = e^{-2\sigma(y_b)}\eta_{\mu\nu}(x).$
After renormalization of
the wave function $H\to e^{\sigma(y_b)}H,$ the Higgs vacuum expectation value is changed to
\begin{equation}
\upsilon_{obs} = e^{-\sigma(y_b)}\upsilon_H\approx
e^{-{\epsilon \zeta^2_\infty\over 4\delta}z_b}\;
   \upsilon_H,
\end{equation}
where the last step holds when $\epsilon\zeta^2_\infty\approx 0.$
This shows that any mass parameter $m$ of the original theory appears to have
\begin{equation}
m_{obs} \approx e^{-{\epsilon \zeta^2_\infty\over 4\delta}z_b}\; m.
\end{equation}
Thus, by changing the relative position $z_b/\delta$ of the 3-brane with respect to
the domain wall boundary, one gets a theory that has exponentially varying masses
relative to the Planck mass.

\section*{Acknowledgement}

This work is supported by the Basic Science Promotion Program of the
Chungbuk National University, BSRI-00-S06.


\end{document}